%% file: main.tex
\documentclass[aps,prl, twocolumn,amsmath,amssymb, superscriptaddress,citeautoscript,floatfix]{revtex4-2}
\usepackage{amsbsy}
\usepackage{wasysym}
\usepackage{bm}
\usepackage{color}
\usepackage{textcomp}
\usepackage{tikz-feynman,contour}
\tikzfeynmanset{compat=1.1.0}
\tikzfeynmanset{/tikzfeynman/momentum/arrow shorten = 0.3}
\tikzfeynmanset{/tikzfeynman/warn luatex = false}
\usepackage{gensymb}
\usepackage[breaklinks,colorlinks = true,linkcolor = red,urlcolor=cyan,citecolor=red]{hyperref}
\usepackage{array}
\usepackage{float}
\usepackage{graphicx}
\usepackage{subfigure}
\usepackage{float}
\usepackage{soul}

\usepackage{comment}
\usepackage{dcolumn,multirow}

\usepackage{xcolor}

\begin{document}

\title{Real-space chirality  from crystalline topological defects in the Kitaev spin liquid}

\author{Fay Borhani}
\affiliation{School of Physics, Georgia Institute of Technology, Atlanta, GA 30332, USA}
\author{Arnab Seth}
\affiliation{School of Physics, Georgia Institute of Technology, Atlanta, GA 30332, USA}
\author{Itamar Kimchi}
\affiliation{School of Physics, Georgia Institute of Technology, Atlanta, GA 30332, USA}

\date{February 16, 2025}

\begin{abstract}
We show that certain crystalline topological defects in the gapless Kitaev honeycomb spin liquid model generate a chirality and Majorana fermion orbital magnetization that depends in a universal manner on their emergent flux. Focusing on 5-7 dislocations as building blocks, consisting of pentagon and heptagon disclinations, we identify the Kitaev bond label configurations that preserve solvability. By computing two formulations of local markers $M(r)$ we find that the 5 and 7 lattice defects generate a real-space contribution to Chern number and an associated Majorana fermion orbital magnetization proportional to $M(r)$. The sign of the $M(r)$ contribution from each 5/7 defect, i.e.\ its $q_M=\pm 1$ chirality, is determined by the defect Frank angle sign $F$ and emergent gauge field flux $W = \pm i$ through the expression $q_M = - i  F W$. Remarkably, though lattice curvature and torsion can interplay with the surrounding gapless background to modify the profile of $M(r)$, its sign $q_M$ is determined locally, implying that crystalline defects in the Kitaev spin liquid  can generate a robust and observable chirality. 
\end{abstract}

\maketitle

\section{\bf {INTRODUCTION}}

In the paradigmatic Kitaev model of a quantum spin liquid (QSL) on the 2D honeycomb lattice, it has been known since Kitaev's original work~\cite{kitaev_anyons_2006} that plaquettes with an odd number of sides carry an imaginary unit of flux $W=\pm i$  of the emergent gauge field, which therefore breaks time-reversal (TR) symmetry.
This feature of imaginary fluxes (equivalently denoted as $\pm\pi/2-$flux) has been demonstrated on various 
lattices decorated with odd-sided plaquettes, where it leads to a chiral QSL.
\cite{yao_exact_2007, dusuel_perturbative_2008, peri_nonabelian_2020} 
Recently this feature was also studied on non-crystalline amorphous lattices 
\cite{cassella_exact_2023, grushin_amorphous_2023}
where it also produces a chiral QSL.
This non-Abelian chiral QSL is a sought-after phase with a quantized thermal Hall conductance which would be quite interesting to produce in Kitaev materials at zero applied magnetic field. However, since Kitaev materials are spin-1/2 magnetic insulators on the honeycomb lattice, far from the amorphous limit, it is not a priori clear how to apply the above results on  chiral QSLs to usual Kitaev materials. 

Here we study the chirality (and associated Majorana orbital magnetization and  thermal Hall conductance) arising from odd sided plaquettes in a complementary viewpoint that is more relevant for the honeycomb lattice
of Kitaev materials. 
Our definition of chirality is the local Chern marker described below, which can be further interpreted as the Majorana orbital magnetization from the Kitaev QSL excitations.
We consider the fundamental honeycomb lattice defect building blocks that if proliferated would lead to an amorphous lattice without translation symmetry,  in particular the honeycomb lattice fundamental 5-7 edge dislocation defect. 
Kitaev materials hosting spin-1/2 moments on the honeycomb lattice necessarily have such honeycomb lattice dislocations.
It is thus necessary to consider the effects of dislocations on the Kitaev QSL in order to correctly analyze experimental data on Kitaev materials. However, the chirality of chiral QSLs (which can be measured by quantized thermal Hall conductance) is a global topological property of the phase, usually associated with a Chern number of gapped Majorana bands in momentum space, and there is not yet a theory describing chirality (equivalently, fermion orbital magnetization) of isolated dislocation defects in Kitaev materials.
In this work we use the Local Marker described below to compute contributions to the chirality in Kitaev systems with isolated dislocations. This computation allows us to extract a measure of a local chirality contributed by each disclination part of a dislocation. We find that the this local chirality  is robustly determined by the defect Frank angle and the emergent flux (Eqn.~\ref{eq_chern charge}). 
This computational extraction of the chirality contributed by each defect building block can enable a microscopic understanding of the role of defects in Kitaev materials, 
and the possible formation of chiral spin liquid (with quantized thermal Hall response) due to the presence of lattice defects.

Dislocations in QSLs have often been studied by focusing on their  braiding statistics 
as ``twist defects'' or ``genons''  \cite{barkeshli_twist_2013, yan_generalized_2023}, which are helpful if the positions of individual dislocations could be controlled independently and coherently. 
Dislocations have also been studied as a source of lattice torsion, equivalently a dipole source of lattice curvature  (since a basic edge dislocation is a dipole of disclinations with opposite Frank angles), including on the honeycomb lattice of graphene \cite{VozmedianoGauge2010}. 
Here we consider a different effect, studying how quenched dislocations can modify the quantum  state in Kitaev materials beyond the conventional non-QSL effects of lattice curvature. This viewpoint was also taken by a prior study of dislocations in the gapped phase of Kitaev's model, created by a large bond strength anisotropy, which protects unpaired Majorana modes on dislocations \cite{petrova_unpaired_2014}. 
The prior study did not consider chirality generation from dislocations, and indeed the lack of global chirality generation is consistent with the strong-anisotropy trivial gap. In contrast, here we consider adding dislocations to the gapless phase of the Kitaev model, where local chirality generation can have important physical effects. 
The gapless Kitaev model, which occurs when the three types of Kitaev exchanges (x,y,z) all have approximately the same energy scale, is the model that is thought to arise (with perturbations) in most Kitaev materials. It is thus necessary to consider possible chirality effects from defects when investigating Kitaev materials. 
Finally we note recent manuscripts studying the Kitaev gapless QSL in hyperbolic space with homogeneous curvature, which was found to either be gapped \cite{mosseri_kitaev_2024} or remain gapless \cite{dusel_chiral_2024, lenggenhager_hyperbolic_2024}
depending on the tiling of the hyperbolic space. 
In contrast, below we find that that a nonzero local gap is generated by the isolated dislocation defects which are relevant to honeycomb Kitaev materials.

Many potential realizations of the Kitaev model Hamiltonian in magnetic insulators have been discussed in the literature, and the field is well reviewed elsewhere 
\cite{savary_quantum_2016, winter_models_2017, hermanns_physics_2018, takagi_concept_2019, trebst_kitaev_2022, rousochatzakis_kitaev_2024}.
In certain materials even weak Kitaev exchanges have been robustly characterized, such as  CrI$_3$ \cite{kim_spin_2024} and YCOB-Cl \cite{seth_disorderinduced_2024}, suggesting the Jackeli-Khaliullin mechanism \cite{jackeli_mott_2009} for generating Kitaev type exchanges can be quite robust. 
Regarding dislocations, as far as we are currently aware, dislocation defects have not been well characterized in any Kitaev material to date. 
However such defects always arise in solids. Characterization of other types of defects have been performed for certain Kitaev materials such H$_3$Li$_2$IrO$_6$ \cite{takayama_hyperhoneycomb_2015, yang_muon_2024, halloran_continuum_2024}. 
Regarding experimental probes of a chiral spin liquid, the theoretical golden standard is quantized $\kappa_{x y}/T$ thermal Hall conductance. Though experimentally complex it has been increasingly used as a probe in Kitaev materials such as RuCl$_3$ and Na$_2$Co$_2$TeO$_6$ 
\cite{kasahara_Majorana_2018, czajka_planar_2023, takeda_planar_2022, hong_phonon_2024}. 
 In our setting the chirality and Majorana fermion orbital magnetization is emergent from non-magnetic lattice defects. Thus if charge fluctuations enable it to couple to real magnetic fields \cite{motrunich_orbital_2006} it will also be observable as a local magnetic moment visible in scanning magnetometry probes \cite{zhou_imaging_2023, persky_magnetic_2022, zhou_sensing_2024}.

The time-reversal-breaking chirality and associated quantized thermal Hall conductance of the chiral QSL has a conventional theoretical description in terms of the Chern number.  
Recall that for 2D electron systems with conserved charge, the Chern number of the gapped electronic band structure determines the quantized Hall conductance. 
For the Kitaev QSL the electrons are insulating and their spins fractionalize into Majorana fermions without U(1) symmetry. Thus (as we discuss below) the Chern number of the  Majorana bands determines the orbital magnetization of Majorana fermions \cite{ceresoli_orbital_2006}. This gives the system's quantized thermal Hall conductance and associated chiral edge modes \cite{kitaev_anyons_2006} for systems with open boundaries. 
However, the presence of defects  prevents the use of the usual momentum-space band-structure methods for determining chirality, such as the topological Chern number. A real space theoretical probe of chirality is thus needed.

Such a set of real-space theoretical probes of chirality have been under development in recent years, due to seminal work by Kitaev \cite{kitaev_anyons_2006} and Bianco and Resta \cite{bianco_mapping_2011}, often going by the name of ``local Chern markers'' $M(r)$ \cite{kitaev_anyons_2006, bianco_mapping_2011, bianco_orbital_2013, dornellas_quantized_2022}. These theoretical probes or ``markers'' are useful in that when they are measured in translationally invariant systems, they exactly reproduce the conventional Chern number. Their definition persists with disorder, for example disorder averaging a fully gapped but inhomogenous system reproduces the Chern number uniformly across all local markers \cite{bianco_mapping_2011, dornellas_quantized_2022, markov_local_2021}. Local markers can also be defined for 2D interacting systems in terms of Green's functions \cite{markov_local_2021} and for topological invariants in other dimensions \cite{hannukainen_local_2022}.
Other methods exist for measuring the global topology of an inhomogenous system, such as Kitaev's ABC construction \cite{kitaev_anyons_2006} and the Bott index \cite{hastings_topological_2011}, which were also used to compute the nonzero chirality of the Kitaev QSL in homogeneous hyperbolic space \cite{dusel_chiral_2024, lenggenhager_hyperbolic_2024}.  However, the local Chern markers $M(r)$ are  advantageous  in giving real space resolution.
Such real space resolution becomes especially important for systems where the gap is only ``local'' \cite{henheik_response_2024}, an issue which arises in the present case of dislocations embedded in an otherwise gapless background,  as we discuss below.

We note \cite{thonhauser_orbital_2005,bianco_orbital_2013} that the local Chern marker $M(r)$ can be physically interpreted as the topological part $\mathcal{M}_\text{topo}$ of the local Majorana orbital magnetization $\mathcal{M}_\text{orbital}(r) =\mathcal{M}_\text{non-topo}(r) 
 + \mathcal{M}_\text{topo}(r)$. 
For the Majorana fermions of the Kitaev model, we show that particle-hole symmetry eliminates the non-topological part $\mathcal{M}_\text{non-topo}$. 
This vanishing of $\mathcal{M}_\text{non-topo}$ persists in the presence of dislocations we study, so that the Majorana fermion orbital magnetization is entirely determined by the topological part i.e.\ the local marker $M(r)$:
\begin{equation}
  \mathcal{M}_\text{orbital}(r) = \mathcal{M}_\text{topo}(r) = \mu_0 M(r).  
\end{equation}
Here $\mu_0$ is the chemical potential measured from the bottom of the band ($\mu_0 = 3 J_K$ in the notation of Eq.~\ref{eq:H}).
We use units such that the fermion gauge charge and prefactor $1/(\hbar c)$ are absorbed into $\mathcal{M}$. Charge fluctuations may couple the Majorana fermion orbital magnetization to the electron magnetization \cite{motrunich_orbital_2006}, but addressing this question here requires a model beyond Eq.~\ref{eq:H} and is beyond the scope of this work.

\section{\bf{RESULTS}}

\textbf{Summary of results}

\begin{figure}[t]
\centering
 \includegraphics[width=0.99\columnwidth]{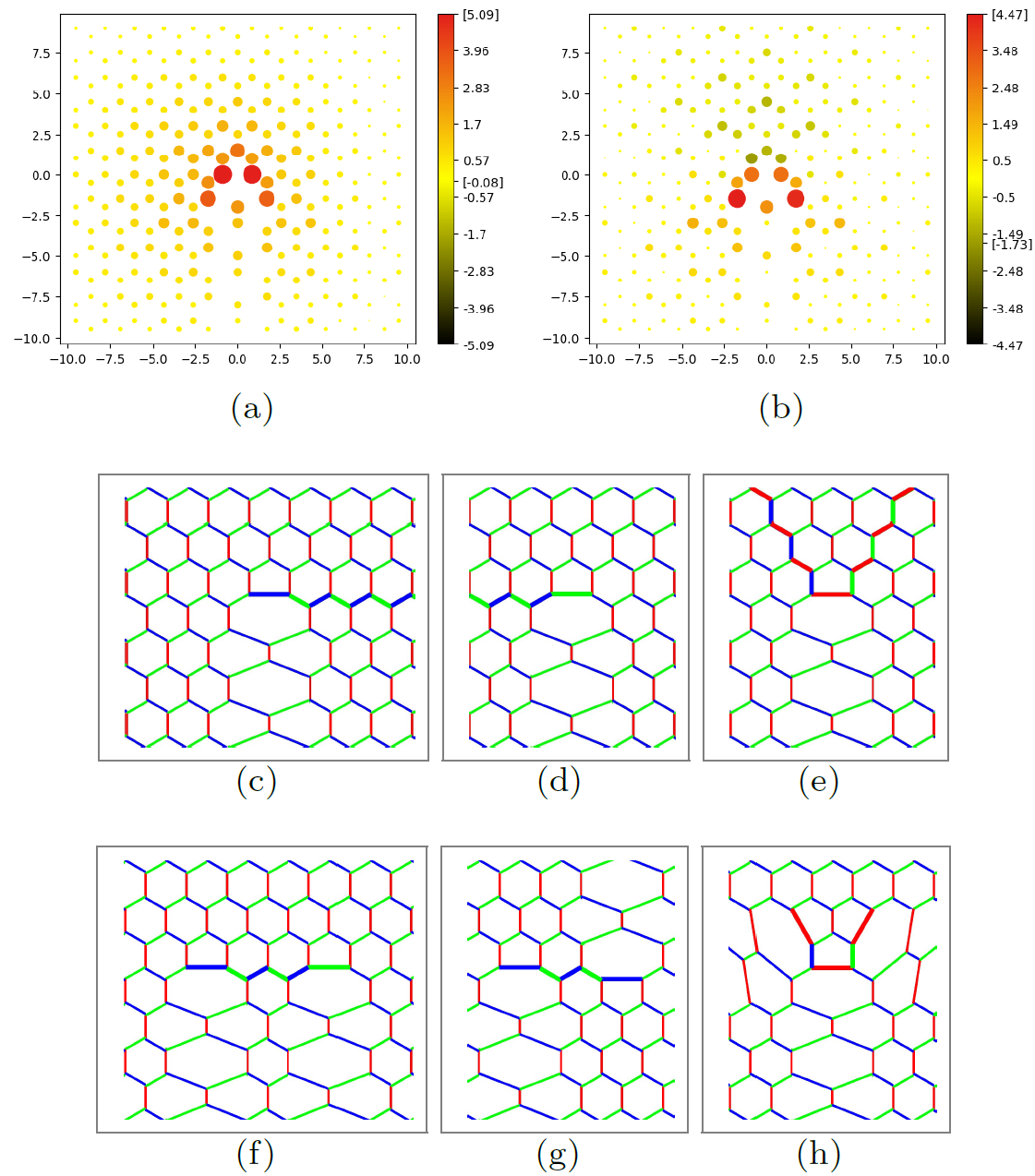}
\caption{{\bf 
5-7 Dislocations in the Kitaev honeycomb model and their contribution to chirality:}
{(a,b)} Pentagon-heptagon (5-7) dislocations, and the $M_1$ local Chern marker pattern they generate, for the case where (5,7) plaquettes carry $W$ fluxes of (+i,-i) (a) and (-i,-i) (b) respectively. 
The local marker patterns can be viewed as arising from a monopole (a) and dipole (b) of chirality charges $q_M$ associated with the defect fluxes. 
$M_1$ on each site is shown as area-$|M_1|$ disk; note $|M_1|$ is largest near the dislocation core and decays away from it. 
Max/min $M_1$ values are shown in brackets on color scale, and are unequal in (b), implying that the $\pm$ charges $q_M$ assigned to (b) should have unequal magnitude. 
{(c,d,e)} Dislocation with Kitaev bond labels that preserve exact solvability.
{(f,g,h)} For multiple dislocations, the flipped-handedness strings of (c,d,e) can be truncated under certain conditions as described in the text.
Note that while each dislocation is drawn on flat space with a ``tail'', this is only a visualization: all lattice bonds have uniform $J_K$ regardless of depicted bond length.
}
\label{fig:one}
\end{figure}

In this work we investigate the chirality contribution of isolated 5-7 dislocations in Kitaev's model. 
By defining Kitaev bond label configurations that preserve exact solvability even in the presence of dislocations (Fig.~\ref{fig:one} c,d,e), we diagonalize the idealized Kitaev model in the presence of various dislocations and study the resulting states. The local Chern marker $M(r)$ is computed to investigate the contributions of such isolated defects to the chirality of the system. We find that these computational results can be naturally interpreted in terms of a ``chirality charge" contributed by each 5/7 disclination building block, whose sign is determined by the disclination Frank angle and the emergent flux.
(Recall that the Frank angle is the topological invariant of a disclination, measuring the angle subtended by a wedge that is required to be added or removed from the pristine lattice to create the disclination.)
In particular, we numerically demonstrate that the sign of the  chirality breaking around each 5 or 7 disclination can be expressed as
\begin{align}
    q_M ~= - i  ~F ~ W
    \label{eq_chern charge}
\end{align}

where $F$ is the sign of the 5 or 7 disclination Frank angle, namely $F=+1$ ($F=-1$) for 5 (7) disclinations respectively. 
The chirality charge $q_M=\pm 1$ gives the sign of the Chern marker distribution and thus also the sign of the fermion orbital magnetization that arises near the disclination. The resulting $M(r)$ profile has the sign $\text{sgn}(M(r)) = q_M$ away from system boundaries, and is localized near the defect.

The possibility of constructing a more complete electrostatics analogy, including magnitudes of combined chirality charges $Q_M$ of multi-defect systems and the resulting spatial profiles and magnitudes of marker and orbital magnetizations $M(r)$, is left for future work. Below we show that such a full theory would necessarily need to incorporate lattice curvature and torsion effects.
However the chirality charge assignment $q_M=\pm 1$ 
already functions similarly to electric charges in electrostatics in creating various multipolar configurations familiar from electrostatics (Fig.~\ref{fig:one} a,b and Fig.~\ref{fig:two_dislocs}).

\begin{figure}[t]
    \centering
 \includegraphics[width=0.99\columnwidth]{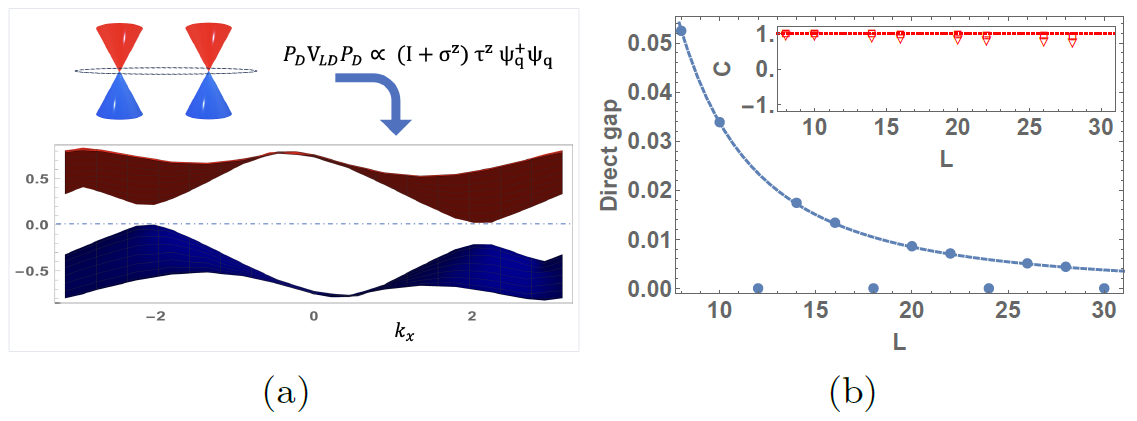}
     \caption{{\bf Chern number $C=1$ generated by periodic array of pseudo-dislocation non-topological local defect and interpretation via perturbation theory:} 
    (a) A periodic array of pseudo-dislocation non-topological local defect $V_\text{LD}$ with zero lattice curvature adds a perturbation that opens a direct gap in each Dirac cone, while also preserving a zero-energy eigenvalue in each cone, thanks to the matrix $(1+\sigma^z)$ that arises when  $V_\text{LD}$ is projected to the low energy Dirac cones. 
    Bottom panel: band structure for an array with $V_\text{LD}$ defect density of 1/16 is here plotted as a function of $k_x$, viewed along $k_y$, to show the nonzero direct gap and barely vanishing indirect gap.
    (b) Computations for periodic array with one $V_\text{LD}$ defect per unit cell with $L^2$ sites, at $\lambda=1$.  The direct gap ($\approx 3.38 L^{-2}$, blue dashed line) remains nonzero for $L\neq 6\mathcal{Z}$ (i.e.\ when the $K_1,K_2$ cones are not both folded to $k=0$), allowing a computation of Chern number of the lower band.  Note that the fluxes corresponding to $V_\text{LD}$ are $[i,-i]$ on [5,7] respectively, so based on the local markers and Eqn.~\ref{eq_chern charge} this is expected to behave as an array of purely positive chirality charges $q_M$. Inset: Indeed, computing the Chern number for the $L\neq 6\mathcal{Z}$ arrays with increasing accuracy (triangle and square symbols), we find that it converges to $C=1$ (dotted line). Thus the local effects of dislocations are sufficient to generate a nonzero Chern number, and can be understood within perturbation theory, with a sign predicted by Eqn.~\ref{eq_chern charge}.}
    \label{fig:local}
\end{figure}

\begin{figure}[t]
 \includegraphics[width=0.99\columnwidth]{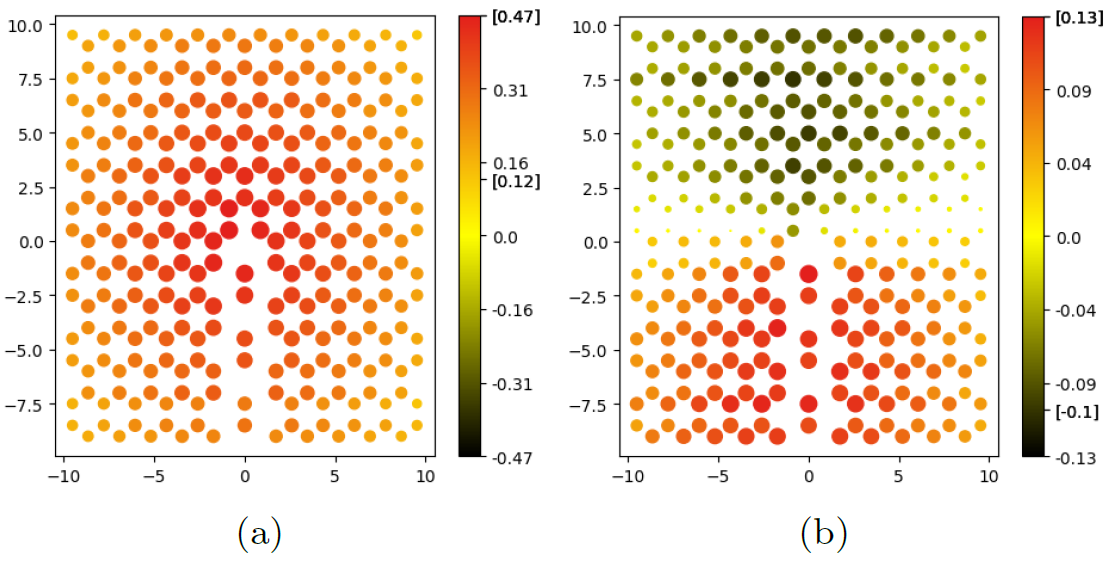}
 \caption{Variant $M_2$ local Chern markers, computed for the same flux patterns on a dislocation as in Fig.~\ref{fig:one}, namely $(+i,-i)$ and $(-i,-i)$ for (a), (b) respectively. Here $M_2(R)$ is plotted for each region $R$ as a disk of area $|M_2|$ at site $r$ below and left of $R$. The profile $M_2(R)$ differs from $M_1(r)$ due to the nearly-gapless background modifying the lattice-scale $M(R)$ manifestation of the physical coarse-grained magnetization $\mathcal{M}(r)$. The chirality  signs Eqn.~\ref{eq_chern charge} are robust.
 }
\label{fig:M2}
\end{figure}

\begin{figure}[t]
 \includegraphics[width=0.99\columnwidth]{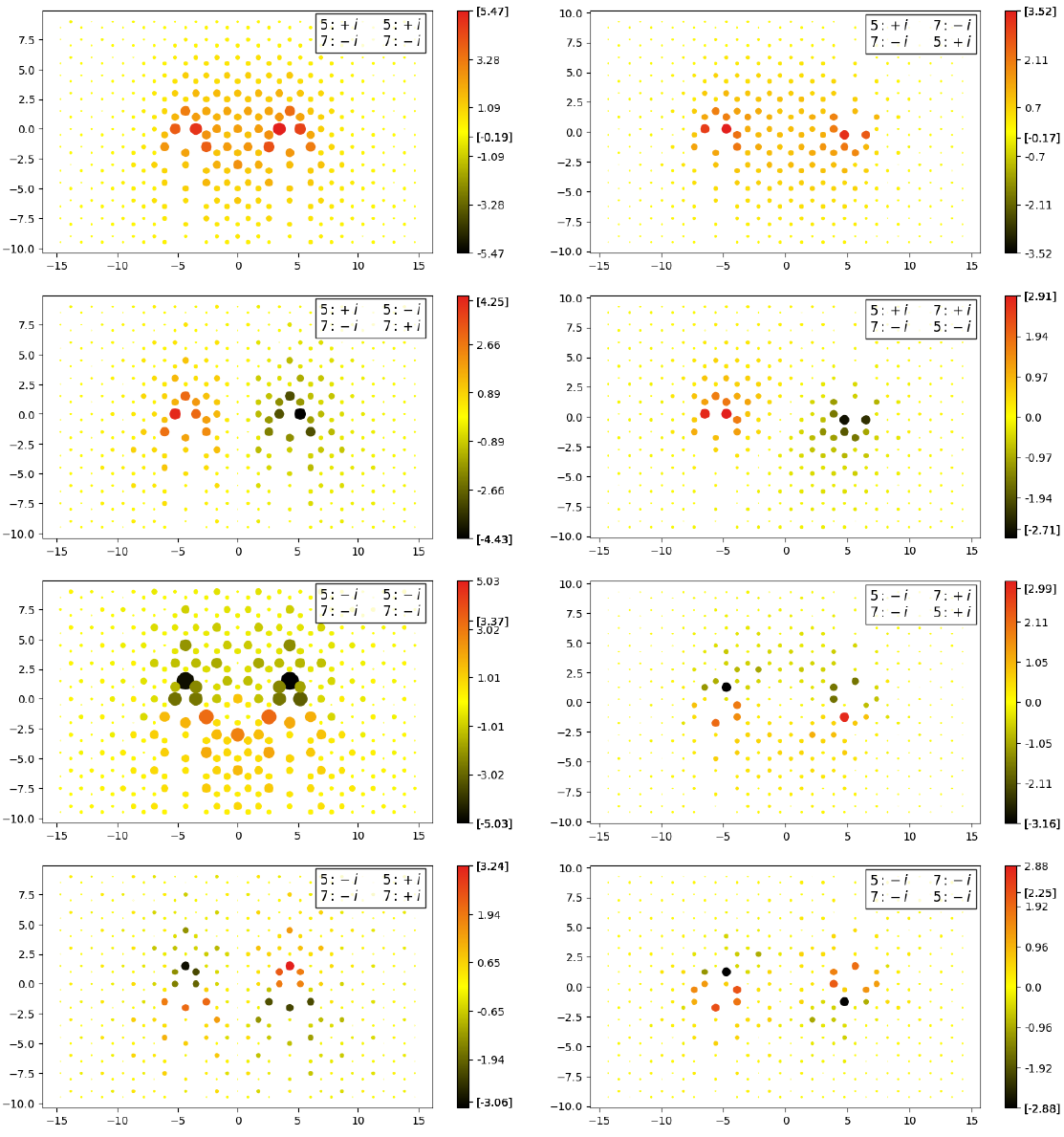}
\caption{For two dislocations, aligned versus antialigned Burgers vectors (left and right columns respectively) yield different $M_1$ local markers configurations, implying that curvature can have beyond-local effects on the local marker. Insets show the fluxes on the 5 and 7 plaquettes corresponding to their spatial positions.  }
\label{fig:two_dislocs}
\end{figure}

A central subtlety of isolated defects is that the system remains nearly gapless. However, we show that the essential  content  of Eq.~ \ref{eq_chern charge} can be extracted by tracking a  local contribution. We confirm this  using  perturbation theory and superlattice Chern number computations on a  pseudo-dislocation non-topological local defect (Fig.~\ref{fig:local}).  We also explore the beyond-local magnitude effects that do depend on curvature and torsion (Figures ~\ref{fig:M2} and ~\ref{fig:two_dislocs}).

\bigskip \textbf{Solvability of Kitaev models with dislocations}

The Kitaev model on the honeycomb lattice \cite{kitaev_anyons_2006} is
\begin{align}
    H=J_K\sum_{\langle ij\rangle}\sigma^{\alpha_{ij}}_i\sigma^{\alpha_{ij}}_{j}
    \label{eq:H}
\end{align}
where  ${\alpha_{ij}} = x,y,z$ denotes the  three axes, shown in Fig.~\ref{fig:one} as  bond colors red, green, blue.
We consider uniform $J_K$ on all bonds even in the presence of defects. (This is true even though our drawings of dislocations on the flat 2D plane involves a sequence of stretched bonds; these bonds are identical to any other bond on the lattice.)

The model and its possible realizations are well reviewed elsewhere \cite{savary_quantum_2016, hermanns_physics_2018, takagi_concept_2019, trebst_kitaev_2022, rousochatzakis_kitaev_2024}. 
Here we briefly review its solution in order to introduce relevant notation. 
The model can be solved via Kitaev's prescription of decomposing  spins in terms of Majorana fermions:
$\sigma^\alpha_i=ib_i^{\alpha}c_i$
where $b_i^\alpha$ and $c_i$ are Majorana fermions sitting at the lattice site $i$. The above Hamiltonian can then be rewritten as
$H={(1/2)}\sum_{i,j}J_K (-i u_{ij}) c_i c_{j}$
where, $u_{ij}=ib^{\alpha_{ij}}_{i}b^{\alpha_{ij}}_{j}$ describes $Z_2$ gauge field residing on the bonds of the honeycomb lattice.  Since $\left[H, u_{i,\alpha}\right]=0$, the gauge field behaves as a static background field, giving rise to exact solvability of the model.  Note that The above Majorana Hamiltonian is equivalent to the complex fermion ($\tilde{c},~\tilde{c}^\dagger$) hopping model $H=J_K\sum_{i,j} i u_{ij}\tilde{c}_i^\dagger \tilde{c}_j$.  For each  closed path  the gauge field gives a flux $W_{\bigcirc}=\prod_{i\alpha\in\circ}(- i u_{i,\alpha})$ (defined with counterclockwise orientation)
which is gauge invariant and conserved. 
As pointed out in Ref.~\cite{petrova_unpaired_2014}, this definition of $W_{\bigcirc}$ coincides with $W_{\bigcirc}$ defined in terms of spin operators around a plaquette with $n$ vertices as a product of $2n$ spin operators: $(\sigma^z_1 \sigma^z_2)(\sigma^x_2 \sigma^x_3)$ for the two sites 1,2 on the $z$ bond, followed by the two sites 2,3 on the $x$ bond, etc, following the bonds counterclockwise around the plaquette.
Note that $W^2_{\bigcirc}=1~(-1)$ on the even (odd) sided plaquettes respectively. This constraint enforces TR symmetric $\pm 1$ flux (equivalently denoted by its phase as $0-$ and $\pi-$flux) in the even sided plaquettes, and TR breaking $\pm i$ flux (equivalently $\pm\pi/2-$flux) on the odd sided plaquettes.

In presence of the dislocation, the honeycomb lattice remains three fold coordinated. Certain 3-coloring configurations that preserve solvability can still be defined, as shown in Fig. \ref{fig:one} (c,d,e), where (c) and the related (d) were also found in Ref \cite{petrova_unpaired_2014} while (e) is new.  Importantly, a dislocation involves an infinite string of sites which have modified Kitaev bond labels relative to the rest of the lattice. These strings also show flipped handedness of $x,y,z$ labels on the bonds around each site. The choice of the string of flipped handedness (SFH) is associated with the choice of color of the 5-7 bond at the dislocation core, as shown in Fig. \ref{fig:one}(c), (d) and (e). String reorientations are related by subsystem symmetry transformations (cycling the bond labels along certain 1D paths). Note that in (c,d) the infinite SFH has a starting point while in (e) it is infinite on both directions. 

For lattices with multiple dislocations, in certain configurations the strings can be chosen to be finite.
A necessary condition is that the dislocations can be grouped into groups of two or three dislocations with each group having its Burgers vector either vanishing (Fig. \ref{fig:one} (g,h)), or taking a value twice of the Burgers vector of a fundamental 5-7 dislocation, i.e.\ two dislocations that have parallel Burgers vectors (Fig. \ref{fig:one} (f)). 
(Recall that the Burgers vector is the topological invariant of a dislocation, measured by the mismatch of the starting and ending point, relative to the pristine lattice, of closed paths that encircle the defect.)
An additional condition is that the dislocation positions align so that the strings can overlap; this can be fairly nontrivial as depicted in (Fig. \ref{fig:one} (h)). 
We note that the finite-SFH requirement of total Burgers vector being zero or twice the vector of a single dislocation is quite general and allows for various coarse grained torsion patterns.

\bigskip \textbf{Single dislocations}

To motivate the computations of chirality and Chern number which we will turn to in the following sections, we begin with a local Chern marker $M_1$ computation  shown in Fig.~\ref{fig:one} (a), (b). 
This local marker is a measure of chirality in that when computed in translationally invariant systems it reproduces the Chern number. (We discuss its definition and application to defects further below.) 
Fig.~\ref{fig:one} (a), (b) show that each dislocation generates a pattern of $M_1$ with substantial nonzero magnitude near the dislocation core. This pattern can be interpreted as arising from a chirality charge associated with each disclination, whose sign depends on the flux in accordance with Eqn.~\ref{eq_chern charge}.

For a single dislocation, the opposite-fluxes configuration is its ground state.
This corresponds to chirality charges of equal sign, giving a ``monopole'' local Chern marker contribution, Fig. \ref{fig:one}. 
The gap from the opposite-fluxes ground state of a dislocation to the excited state of uniform-fluxes configuration is approximately $\Delta_\text{uniform-fluxes} \approx 0.038 J_K$.

Since odd-sided plaquettes also carry curvature, a natural question is whether the role of this curvature in setting the chirality is significant or not. A second question is how to relate the local Chern markers $M_1$ and $M_2$ (discussed below) to the traditional Chern number defined for periodic lattices, given that a unit cell with nonzero net Burgers vector cannot have the 2D translation symmetry necessary for defining the conventional Chern number.

\bigskip \textbf{Isolating  contributions to Chern number from time-reversal breaking  local perturbations}

In this section we address both of these questions by defining a pseudo-dislocation non-topological local defect $V_{LD}$ that does not carry curvature. We show that its 2D arrays produce nonzero Chern number with an appropriate sign corresponding to Eqn.~\ref{eq_chern charge}. We also show how this can be understood within perturbation theory on a dislocation. The results show that the sign of chirality generation arises from purely local perturbation  $V_{LD}$. In the next section we will argue that, in contrast, the magnitude does involve beyond-local effects related to the pattern of curvatures.

Let us begin by defining a
perturbative point of view to describe a dislocation. We note that 
dislocations, being inherently non-local, cannot
be described as a local perturbation to the pristine honeycomb lattice. 
However, we show that their effects on the magnetic sector can be closely mimicked by introducing a single second nearest neighbor bond violating the bipartiteness of the lattice. The Kitaev model with such a local pseudo-dislocation is obtained by adding the following perturbation:
\begin{align}
    V_{\text{LD}}=i\lambda~ c_{{ R},\mu}c_{{ R+d_1},\mu} \rightarrow i\lambda\left(\tilde{c}^\dagger_{R,\mu}\tilde{c}_{R+d_1,\mu}-\tilde{c}^\dagger_{R+d_1,\mu}\tilde{c}_{R,\mu}\right) 
\end{align}
where  ${ R}$ is the position of the {pseudo-dislocation}, ${ d_1}$ is a lattice vector, $\mu$ is one of the two sublattice indices, and $\lambda$ is the hopping strength of the second neighbor bond, which we treat as a perturbative parameter. We also indicate the equivalent hopping in terms of complex fermions. The particular hopping chosen above leads to TR breaking opposite fluxes on the 5- and 7-sided plaquettes around that bond, analogous to Fig.~\ref{fig:one} (a).

The above perturbation, being local,  is not diagonal in the momentum basis. To the leading order in $\lambda$, it generates  matrix elements between two sublattices of the Dirac cones. 
We use a notation such that the low energy theory of the Majorana fermions is rewritten as complex Dirac fermions. 
This double counts states, which is easily fixed by a  factor of 2 when computing total energies, i.e.\ the complex fermions have hopping $J_K$. 
This gives a Hamiltonian which when projected to the Dirac cones reads $ P_D H P_D =  v_F\sum_{ q} {\psi}^{\dagger}_{ q}\left(q_x\sigma^x+q_y\tau^z\sigma^y\right){\psi}_{ q}$. Here $v_F=3J_K/2$, $\sigma$ and $\tau$ denote the  sublattices and valleys, respectively, $P_D$ is the projector onto the low energy Dirac cones, and $\psi_{ q}=\left(c_{{ K_1+q},A}~c_{{ K_1+q},B}~c_{{ K_2+q},A}~c_{{ K_2+q},B}\right)^T$ with ${ K_1}$ and ${ K_2}$ denoting the two Dirac cones. In this notation, the local pseudo-dislocation perturbation leads to an effective low-energy Hamiltonian of the  form 
\begin{align}
    P_D V_\text{LD}P_D = 
   \frac{\sqrt{3}\lambda}{2\mathcal{N}}\psi_{ q}^\dagger(I+\sigma^z)\tau^z\psi_{ q}
\end{align}
where $\mathcal{N}$ denotes the total number of unit cells. The first term breaks the inversion symmetry  giving rise to tilt in the band structure (see Fig. \ref{fig:local}). The second term, $\sigma^z \tau^z$, breaks time-reversal and generates a topological mass term for the Dirac cones. This is the central feature of the magnetic physics in presence of a true 5-7 dislocation. Both tilt and splitting  grow linearly with $\lambda$. However, we note that since there is no non-bipartite coupling for the sublattice not chosen by $\mu$, this sublattice exactly preserves the zero energy states, leading to zero gap in the many particle eigenstates. 

The nonzero Chern number that results from this topological mass term is seen explicitly in our numerical studies of $V_\text{LD}$ periodic arrays,  Fig. \ref{fig:local}. Note the tilt between the two Dirac cones, and the single particle gap which remains nonzero throughout  the Brillouin zone, despite the zero energy eigenvalues, thus enabling the rigorous definition of Chern number, which is found to be $C=+1$. This array corresponds to an array of the $(+i,-i)$ dislocation of Fig.~\ref{fig:one}(a), restricted to only local effects of the dislocation with no curvature. Thus the positive $M_1$ generated by the dislocation of Fig.~\ref{fig:one}(a) can be fully captured by the purely local effects of the dislocation, and it indeed results in nonzero positive Chern number. 

We further note that  in presence of pseudo-dislocations,  particle-hole symmetry $E_k = - E_{-k}$ is preserved. As described in the Methods section, this particle-hole symmetry in the Majorana fermion description of the Kitaev model enforces the vanishing of $\mathcal{M_\text{non-topo}}$.  
We also find numerically that $\mathcal{M_\text{non-topo}}$ vanishes in configurations with true dislocations up to corrections of order  $10^{-6}$ arising from open boundaries.
Thus the magnetization is entirely controlled by the topological part $\mathcal{M_\text{topo}} = \mu_0 M(r)$ which can be formulated in terms of the local marker. With this result in mind, we now turn to the local marker computations.

\bigskip \textbf{Computation of local Chern marker with a nearly-gapless background}

In general, a local Chern marker $M(r)$ is a real space decomposition of Chern number $C$ as $C=\frac{1}{N}\sum_{r}M(r)$ with $N$ the number of sites in the sum. As discussed above, it also gives the topological contribution to  orbital magnetization .
 The local marker is a lattice scale manifestation of the physically observable magnetization $\mathcal{M}(r)$. The latter can be viewed as a coarse graining of a lattice-scale marker.
This distinction between variant formulations of local markers, which represent variant lattice scale representations of coarse grained $\mathcal{M}(r)$, becomes important in a system with a nearly gapless background as in the present case.

There are presently multiple competing formulations of local Chern markers, which  agree for a uniformly gapped bulk system, but can differ in the presence of inhomogeneities and boundaries, which are necessarily involved for certain configurations of dislocations. \cite{bianco_mapping_2011,dornellas_quantized_2022}
In Figs.~\ref{fig:one} and \ref{fig:two_dislocs} we plot one variant of the local Chern marker, which we call $M_1$, which we introduce  here through a modification of the  $M_{2011}$ marker of Ref \cite{bianco_mapping_2011}  that avoids unphysical edge-mode contributions from open boundaries in the presence of the nearly gapless background.
 $M_{2011}$ is defined as $ M_{2011}(r) = 4 \pi  \text{Im}  \langle r| P x P y P |r\rangle 
$ 
where $P$ is projection to occupied states. 

The issue with open boundaries can be seen more clearly by rewriting $M_{2011}$ following Ref  \cite{dornellas_quantized_2022} in terms of the object $C(r,R)$, which we dub a``contribution map'',
\begin{align}
C(r,R) =  4 \pi \text{Im}  \langle r| P \theta_{x;R} P \theta_{y;R} P |r\rangle .
\label{eq_contribution map}
\end{align}
Here  $\theta_{y;R}$ is the step function on the upper half plane above a point $R$, namely $\theta_{y;R}$ is 1 above $R$ and 0 below; $\theta_{x;R}$ follows the analogous definition. 
The step functions $\theta_{x,y;R}$ have a physical interpretation as the electric potential and the orthogonal flow of charges that together define Hall conductance.  \cite{dornellas_quantized_2022}

The marker $M_{2011}$ is a sum over all $R$ as 
$
M_{2011}(r) = \sum_{R} C(r,R)
$. 
However, in the present case of a local gap generated by a lattice defect on top of a nearly gapless background, the positions $R$ near an open boundary sample the edge mode contribution, with an opposite sign, and are not physical. (This is an issue for numerical computations of a dislocation defect that require reasonable sized finite systems; it is of course not an issue in the definition of local markers 
for a real quantum material.)
Indeed the total sum $\sum_r M_{2011}(r) = \sum_{r,R} C(r,R) = 0$ identically vanishes.

We extend $M_{2011}$ by defining a  marker $M_1$ in the following way, which also enables a re-writing in terms of $C(r,R)$ as follows,
\begin{align}
M_1(r) = 4 \pi  \text{Im}  \langle r| P x_b P y_b P |r\rangle
= \sum_{R \in \text{bulk}} C(r,R)
\label{M_1}
\end{align}
where $x_b = L_b^{-1} \sum_{R \in  \text{bulk}} \theta_{x;R}$ and similarly for $y_b$. 
The operators $x_b,y_b$ are position operators $x,y$ whose eigenvalues saturate when approaching the boundaries. This determines the normalization $L_b^{-1}$ of the $\sum_R$ sum. In practice the $x_b, y_b$ terms are simpler to compute while the $C(r,R)$ sum is more physically transparent.  The particular definition of ``bulk'' used in the computations is described in the Methods section. 
Such separations between bulk and boundary have been previously discussed in the literature for local marker and interacting local markers  \cite{dornellas_quantized_2022, markov_local_2021}.

In order to identify the limitations of the local marker lattice-scale realization of coarse grained Majorana orbital magnetization, for the present case with inhomogeneities and a nearly gapless background, we compare our computation of the marker $M_1$ to an alternative local marker formulation, $M_2$. Here $M_2$ is the $M_{2022}$ local marker of Ref \cite{dornellas_quantized_2022} but with a different parameter for bulk-boundary separation, needed due to the nearly gapless background, as described in the Methods section. Thus
\begin{align}
M_{2}(R) = \sum_{r \in  \text{bulk}} C(r,R)
\end{align} 
where the sum over sites $r$ is restricted to sites in the bulk.
Choosing this restriction enables the relation 
\begin{align}
\sum_{r \in \text{bulk} } M_1(r)=\sum_{R \in \text{bulk}}M_2(R) =  \sum_{r, R \in \text{bulk}}  C(r,R) .
\end{align} 
The two variant markers produce the same coarse grained orbital magnetization. 

The $M_2$ results for a single dislocation are shown in Fig.~\ref{fig:M2}.
It is evident that $M_2$ is less sharply peaked (i.e.\ not as local) and involves more coarse-graining compared to $M_1$.  
 However, though the peak profile is modified compared to $M_1$, the monopole and dipole sign structure remains identical between $M_1$ and $M_2$.

To identify the role of the nearly gapless background with the topological lattice defects, we proceed to consider pairs of nearby dislocations and compute the local marker $M_1$ they generate. Results are shown in Fig.~\ref{fig:two_dislocs}. A few things become evident. First, the magnitude of the local marker generated by each disclination depends sensitively on the curvature profile of nearby space. Recall that each dislocation can be viewed as a dipole of opposite curvatures. Whether the dislocations have their Burgers vectors, and correspondingly their curvature dipole vectors, aligned (Fig.~\ref{fig:two_dislocs} left column) or anti-aligned (right column) greatly impacts the magnitude and profile of the local markers. Second, and in contrast, the pattern of signs of local markers remains invariant, and obeys Eqn.~\ref{eq_chern charge}.

\section{\bf {DISCUSSION}}

In this work we introduce the notion of chirality charge $q_M$ to capture the results of local marker computations near 5-7 dislocations of the Kitaev model. We find that each disclination contributes to the local marker with a sign that is determined by its Frank angle (curvature) and by its flux. This result holds unambiguously even across different formulations of the local markers $M_1,M_2$, and can also be seen to arise perturbatively through local effects as well as manifesting in usual Chern numbers of periodic arrays.  

We also find unambiguous evidence that lattice curvature can impact the magnitude of the Majorana orbital magnetization resulting from the same chirality charge $q_M$, as in the nonequal dipole pattern of Fig.~\ref{fig:one}(b) and the patterns of Fig.~\ref{fig:two_dislocs}.
These results also show that the clean limit's gapless Majorana fermions  are important for a quantitative understanding of the generation of chirality that results from introducing a topological defect. Heuristically, the gapless fermions can carry information about lattice curvature across large length scales.  Thus it is clear that any quantitative theory of how curvature controls $M(r)$ would depend sensitively on the nature of the surrounding background: gapless, topologically gapped or trivially gapped through anisotropy.  

The gapless scenario poses interesting questions.  Due to the gapless background, any ``edge states'' would be extended. It will be interesting to consider how chirality is generated when defects occur at small but finite  density \cite{grushin_amorphous_2023, a.sethetal_chiral_2025}. It will also be interesting to consider realizations of Kitaev models and the appearance and properties of dislocations, disclinations and other topological defects in those models.

\section{\bf{METHODS}}

{\bf Local marker:} $M_1$ marker was computed with a sum over ``bulk" sites defined as being away from the open boundary at a real space distance  $b$, with $b=5$.
Real space distances are defined with a honeycomb lattice nearest neighbor vector of length 1. Due to the emergent Lorentz symmetry of Dirac cones at low energies, relevant to the Majorana fermions of the gapless Kitaev model, real space distance was found to be a better measure compared to numbers of unit cells. 
$M_2$ marker was computed with open boundaries and with a sum over ``interior bulk" sites defined as being within a real space region centered at the origin, with square aspect ratio and real space size $c \times c$, with $c=24$.  This defines the marker $M_2$ within a smaller square region centered at the origin. Computations were performed with the following system sizes. Here system sizes are given by the number of sites along the zig-zag edge and armchair edge, i.e.\ twice the number of a four site unit cell, such that the total number of sites is $ L_x   L_y$, and the real space extent is $\sqrt{3} L_x/2 \times 3 L_y/2$.
Fig 1 (a,b): 
 $(L_x,L_y)=(40,36)$. 
Fig 3: $(L_x,L_y)=(40,36)$.
Fig 4: $(L_x,L_y)=(56,36)$.
Fig 2 Berry curvature integration was performed to obtain Chern number using 72$\times$72 and 144$\times$144 mesh grids denoted by triangle and square symbols respectively.
Diagonalization of the Majorana spectrum was performed using the python PyBinding package \cite{moldovan_pybinding_2020}. 

{\bf Non-topological Magnetization:} For a translation invariant system, the non-topological magnetization can be computed from the band structure using the following equation~\cite{thonhauser_orbital_2005, ceresoli_orbital_2006}:
\begin{align}
    \mathcal{M}_\text{non-topo}
    &=\frac{1}{2}\text{Im}\sum_n\int \frac{d^2k}{(2\pi)^2}\nonumber\\
    &~~~~\langle \partial_k u_{nk}\mid \times \left(H_k + E_{nk}\right)   \mid \partial_k u_{nk}\rangle
\end{align}
where $\mid u_{nk}\rangle$ are the  filled bands with energy $E_{nk}$. For a particle-hole symmetric Hamiltonian, the two contributions $H_k$ and $E_{nk}$ cancel each other. For example, this is the case for the Haldane model with pure imaginary  second neighbor hopping. \cite{thonhauser_orbital_2005, ceresoli_orbital_2006}
In the case of the Majorana Hamiltonian with non-perturbative defects
the band structure formula is no longer applicable. The local formula for the magnetization density is given by~\cite{bianco_orbital_2013},
\begin{align}
    \mathcal{M}_\text{non-topo}(r)=\frac{1}{A}\text{Im}\langle r\mid  P xHyP\mid r\rangle
\end{align}
where $A$ is the area of the system. However, even in presence of the dislocation the particle-hole symmetry is preserved. The non-topological component of orbital magnetization is vanishingly small and consistent with zero: for a 48x48 system with open boundaries our numerical computations find $2 * 10^{-6}$ which is consistent with boundary effects and a zero answer. Hence the local marker, which produces the topological contribution, is sufficient for identifying the magnetization.

\section{\bf{ACKNOWLEDGEMENTS}}
This work was supported by the U.S. Department of Energy, Office of Science, Basic Energy Sciences, under
Award Number DE-SC0025478.


\section{\bf{DATA AVAILABILITY}}
The datasets generated and/or analyzed during the current study will be available at the Georgia Tech Digital Repository following publication.

\section{\bf{REFERENCES}}
\input{main.bbl}

\section{\bf{Author contributions}}
F.B., A.S., and I.K. all contributed to designing the research, writing the code, performing the research, and writing the manuscript.

\section{\bf{Competing Interests}}
The authors declare no competing interests.

\end{document}

%% file: main.bbl
%